\documentstyle[prl,aps,twocolumn]{revtex}
\large  

\input epsf
\begin{document}
\draft \twocolumn[\hsize\textwidth
\columnwidth\hsize\csname
@twocolumnfalse\endcsname
\title{QUANTUM FLUCTUATIONS OF CHARGE AND PHASE TRANSITIONS OF A LARGE 
COULOMB-BLOCKADED QUANTUM DOT}
\author{Eugene B. Kolomeisky$^{*}$, Robert M. Konik$^{\dagger}$, 
and Xiaoya Qi$^\#$} 

\address {Department of Physics, University of Virginia, P. O. Box 400714,
Charlottesville, VA 22904-4714}
\maketitle
\begin{abstract}
We analyze ground-state properties of a large gated quantum dot coupled via a 
quantum point contact to a reservoir of one-dimensional spinless electrons 
whose interactions are characterized by the Luttinger liquid parameter $g$.  
We find that the classical step-like dependence of the average number of 
electrons on the dot $n$ as a function of the gate voltage $n_{x}$
is preserved under certain conditions, {\it the presence of
quantum fluctuations notwithstanding}.  We point out that in its low-energy 
limit the problem is dual to that of a single-junction SQUID 
with Caldeira-Leggett dissipative coupling, and analogous to the classical
problem of multilayer adsorption, and so is related to the classical 
statistical-mechanical problem of a one-dimensional Ising model with exchange 
interactions decaying as the inverse-square of distance.  This Ising 
universality class further subdivides into what we term
(i) the Kondo/Ising class and (ii) the tricritical class.  (i)  For
systems of the Kondo/Ising class, the $n(n_{x})$ dependence is  
always continuous for $g \ge 1$, while for $g < 1$ (repulsive electrons in the 
constriction) the $n(n_{x})$ dependence is continuous for sufficiently
open dots, while taking the form of a modified staircase for dots sufficiently
isolated from the reservoir.  At the phase transition 
between the two regimes the magnitude of the dot population jump is only 
determined by the properties of the reservoir and given by $g^{1/2}$.  
(ii)  For systems in the tricritical class we find in addition an intermediate 
regime where the dot population jumps from near integer value to a region of 
stable population centered about a half-integer value.  In particular,
this tricritical behaviour (together with the two dependencies already seen in 
the Kondo/Ising class) is realized for non-interacting electrons, $g=1$.

\end{abstract}
\vspace{2mm} 
]

\narrowtext

\section{INTRODUCTION}

Quantum dots are nanometer-scale structures hosting a few to a few thousand 
electrons \cite{Jacak} and coming in a variety of forms.  Quantum 
dots can be realized as small metallic islands \cite{ralph},
gated two-dimensional electron gases in semiconductor heterostructures
\cite{scqdots1,scqdots2,scqdots3}, and most recently as short segments of 
metallic carbon nanotubes \cite{tubes}.  The latter two realizations are of 
particular interest as they exhibit a remarkable tunability.  Through the 
application of metallic gates upon either the heterostructures or upon the 
nanotubes, the chemical potential together with the strength of any putative
dot-lead coupling can be tuned.  The associated flexibility afforded to the 
experimentalist has helped to make quantum dots and their related phenomena 
an active area of research.

One such phenomenon that has been closely studied, both experimentally
and theoretically, is Coulomb blockade.  Coulomb blockade in its simplest 
form is a fascinating effect arising from the discrete nature of electrons 
combined with their mutual repulsion \cite{Averin&Likharev}.  The origin of 
the effect is entirely classical.  Imagine charging a dot of capacitance $C$ 
biased by a gate voltage $V_{G}$.  The 
classical electrostatic energy of the dot, $E_{cl} = Q^{2}/(2C) - QV_{G}$, is 
minimized if the dot charge $Q$ equals $CV_{G}$.  But the dot charge can only 
change in the increments of the electron charge $e$, and the dimensionless 
gate voltage $n_{x} = CV_{G}/e$ is generally not an integer \cite{next}. 
Typically there is only one integer closest to $n_{x}$ which determines the 
number of electrons on the dot to be $n = Q/e = [n_{x}]$.   
Half-integer values of $n_{x}$ play a special role because now {\it two} 
states of the dot with $n = n_{x} \pm 1/2$ electrons have the same energy.  
Thus as $n_{x}$ is varied between two nearest half-integers, the dot charge 
is fixed and quantized (the Coulomb blockade) but at half-integer values
of $n_{x}$ the blockade is momentarily lifted, and the dot charge changes 
discontinuously by $e$, leading to the celebrated Coulomb staircase 
\cite{Averin&Likharev}.  

This simple classical picture is modified by quantum charge fluctuations 
arising from the coupling to a reservoir of electrons.  
Quantum mechanical tunneling of charge between connecting leads and the dot 
has been argued to smear the Coulomb staircase even at zero temperature
\cite{glaz}.  Although originally treated perturbatively \cite{glaz}, the 
smearing finds its strongest support in a connection of the quantum dot 
problem to a highly anisotropic Kondo problem \cite{Matveev91}.  In the 
limit of low transparency, i.e. a small tunneling matrix element, and of 
large Coulomb repulsion, the charge dynamics on the dot can be approximated 
by that of a two-level system, with the levels corresponding to charge states 
of the dot differing by unit charge \cite{Matveev91}.  Changing the 
charge on the dot then becomes akin to flipping an impurity spin-1/2.   In 
Kondo language the net charge on the dot parallels the magnetization of the 
impurity spin while the deviation of the gate voltage from the half-integer 
$n_x$ is akin to a Zeeman field applied to the Kondo spin.  For a 
single-channel Kondo problem, the spin susceptibility is well-known to be 
finite at zero applied field.  This correspondingly necessitates the smearing 
of any discrete charge jump with the variation of the gate voltage.

The connection to a Kondo problem applies in the weak-tunneling limit.
The opposite high-transparency limit was treated by both Flensberg 
\cite{Flensberg} and Matveev \cite{Matveev95}.
(For a comprehensive review of 
both cases, see \cite{Aleiner}; for more recent work on this limit
see \cite{LeHur}.)  On the basis of this work it was possible 
to conjecture that the smeared Coulomb staircase of the weak-tunneling limit
smoothly evolves as tunneling is increased into a strictly linear function of 
the gate voltage.  These predictions \cite{Matveev95} found partial 
confirmation in recent experimental work\cite{Berman}.  The agreement between 
theory and experiment was limited by finite temperature effects, with
the temperature of the quantum dot being far in excess of the putative Kondo 
temperature.
                  
In this paper the role of quantum fluctuations is re-examined for a large 
quantum dot - the dot whose charging energy $e^{2}/(2C)$ is substantially 
larger than the distance between the energy levels on the dot.  We show 
that as the strength of the tunneling is tuned, changing the dot from open to 
closed, a modified Coulomb staircase 
reappears under certain circumstance with tunneling non-zero.  
In particular we find that quantum fluctuations need not 
destroy the Coulomb blockade.  We in part arrive at our general conclusion by 
relating the zero-temperature dynamics of spinless electrons to 
the classical statistical mechanics of a one-dimensional Ising model with 
ferromagnetic interactions decaying as the inverse-square of distance.  Within 
the Ising universality class, we find that all experimental systems must be 
either of the Kondo subclass  (to which the connection to the Kondo system 
given in \cite{Matveev91} applies) or of a novel tricritical type.  

The organization of the paper is as follows.  In Section II we set up the
problem in general terms, showing how one understands the problem as 
belonging to the Ising universality class.  We demonstrate how the two 
subclasses we identify can be differentiated in terms of the form of their 
effective potentials.  In Sections III and IV we examine these two subclasses, 
the Kondo and tricritical subclass, in detail.  In Section V we discuss our 
results.

\section{FORMULATION OF THE PROBLEM}

The experimental setup we have in mind is a gated quantum dot 
placed in a large magnetic field (so that the electrons can be considered 
spinless) coupled to a single reservoir (lead) via a narrow constriction - the 
quantum point contact \cite{Matveev95}.  A related setup has been 
experimentally realized \cite{scqdots1,scqdots2,Berman}.  

Regardless of the nature of the reservoir, the problem is effectively 
one-dimensional \cite{Flensberg,Matveev95}.  If the reservoir is 
high-dimensional, the point nature of the contact allows one to dimensionally 
reduce the electrons in the lead through a partial wave expansion, 
keeping only the s-waves, to a one-dimensional problem in much the same way 
the problem of Kondo impurities embedded in bulk metals can be reduced to a 
one-dimensional problem \cite{Chamon}.  

The most general phenomenological Euclidean action for the problem is
\begin{equation}
\label{Action}
A = {\pi \hbar \over g} \int \limits _{- \Lambda}^{\Lambda}{d\omega \over 2\pi}|\omega|
|\phi(\omega)|^{2} + \int d\tau E_{0}(\phi, n_{x}),
\end{equation}
where
\begin{equation}
\label{Clenergy}
E_{0}(\phi, n_{x}) = U(\phi) + {e^{2}(\phi - n_{x})^{2}\over 2C}.
\end{equation}
Here $\tau$ is the imaginary time, and $\phi(\tau)$ is a fluctuating electron 
number field - the expectation value of $\phi(\tau)$ gives the average 
number of electrons on the dot, $n = <\phi>$.  

The first kinetic term in the above action arises in several steps.  In the 
limit of large dot size the time scale by which electrons scattered through 
the point contact return to the scattering point is much larger than any 
other time scale in the problem \cite{Flensberg,Matveev95}.  The point 
contact then connects in effect two reservoirs of electrons.  Upon 
dimensionally reducing the electrons in the higher-dimensional 
leads, one 
obtains a kinetic term appropriate to one-dimensional  
electrons.  Bosonization of the electrons leads to a kinetic term of the form
$$
\int dx d{\tau} (\partial^\mu\phi\partial_\mu\phi).
$$
While the term in the action describing tunneling between the
dot and the lead is naturally restricted to a spatial point [first term of 
(\ref{Clenergy})], it was shown \cite{Matveev95} that the Coulomb charging 
[second term of (\ref{Clenergy})] can also
be represented as acting at a single spatial point.
As these two interaction terms are restricted solely to a single point
we can integrate out the bosonic degrees of freedom away
from the point contact.  Doing so reproduces the first term of (\ref{Action}) 
with $g = 1$ which corresponds to noninteracting electrons in the 
constriction \cite{Kane&Fisher}. 
The frequency cutoff,  $\Lambda$, 
that the kinetic term is equipped with is of 
order the ratio of the sound velocity to the electron density (both in the 
vicinity of point contact).

In (\ref{Action}) we look at a somewhat more general problem by allowing
the one-dimensional electrons to interact through the inclusion
of the dimensionless Luttinger liquid parameter  $g$ which is generally 
different from unity.  This case can be realized if 
both the reservoir and the dot are one-dimensional systems (having common $g$) 
such as quantum wires, metallic carbon nanotubes or fractional quantum Hall 
edges \cite{Geller}. 

The perturbing energy $E_{0}(\phi, n_{x})$, as previously indicated, consists 
of two terms acting at a single point: a term by which electrons 
tunnel through 
the point contact and a term expressing the charging energy of the dot.
The former may generally be represented as a term both
periodic and even in $\phi$, 
$$
U(\phi + 1) = U(\phi); ~~~U(\phi ) = U(-\phi ).
$$
The latter, charging, term takes the natural form
${e^2\over 2C}(\phi - n_{x})^2$.
The  function $U(\phi)$ having its minima at integer and maxima at 
half-integer $\phi$ reflects the preferred tendency for the dot to host an 
integer number of electrons.    The classical limit \cite{Averin&Likharev} is 
recovered by ``closing'' the dot, i.e. by increasing the amplitude of $U$ to 
infinity.  On the other hand, the regime of perfect transmission is reached 
by ``opening'' the dot, i.e. by decreasing the amplitude of $U$ to zero.  While
in the limit of almost perfect transmission, the form 
$$U(\phi) \sim \cos2 \pi \phi$$ 
was derived in \cite{Matveev95} through bosonization,
the precise functional form of $U(\phi)$ should depend on the 
details of the constriction.

The above action finds applications beyond that of quantum dots.
The inspection of (\ref{Action}) and (\ref{Clenergy}) reveals that in the 
low-energy limit the Coulomb blockade problem is {\it dual} to that of 
a single-junction SQUID \cite{Likharev} with a Caldeira-Leggett 
\cite{Caldeira&Leggett} dissipative coupling to the environment.  Here the 
magnetic flux through the SQUID loop is dual to the charge of the dot $Q$ 
(the flux quantum $\Phi_{0}$ is dual to the electron charge $e$), and so all 
our conclusions will have their dual counterparts.  This example of 
electric-magnetic duality is discussed in further detail by 
Likharev \cite{Likharev}.

Alternatively by viewing the imaginary time coordinate $\tau$ in 
(\ref{Action}) as a fictitious space direction we arrive at an effective 
Hamiltonian defining a classical statistical mechanics problem \cite{Kogut}.  
The above action can be then recognized as describing a one-dimensional 
version of the problem of multilayer adsorption 
\cite{Weeks,Huse,Weichman&Prasad}.  
In this problem a film in contact with a liquid is adsorbed onto an attractive 
substrate.  The film thickness (controlled by external pressure) is finite as 
the free energy of the material of the film exceeds that of the liquid.  For 
thick films the competition between the substrate attraction and the confining 
effect of external pressure can be modeled by a parabola centered near optimal 
film thickness \cite{Weeks}.  For a {\it crystalline} film there is also an 
additional periodic contribution into the free energy which oscillates with 
the number of adsorbed atomic layers.  If the field $\phi$ is regarded as a 
number of adsorbed atomic layers, then (\ref{Clenergy}) can be identified with 
the uniform part of the film free energy.  The $|\omega|$ dependence of the 
first term of (\ref{Action}) translates into a $1/\tau^{2}$ repulsion occuring 
between two parts of the film having different heights and separated by 
distance $\tau$.  The dependence of the film thickness upon pressure is 
analogous to $n(n_{x})$.

In what follows we will be interested in the $n(n_{x})$ dependence.  
This quantity has been shown to be directly accessible experimentally. 
Through fabricating two quantum dots in close proximity, the authors of
\cite{Berman} were able to measure the charge on one dot by monitoring the 
current through the other.  The $n(n_{x})$ dependence has an inversion 
symmetry about every half-integer point $n = n_{x}$, and the whole dependence 
can be reconstructed from any segment of $n(n_{x})$ of unit length.  Thus 
without the loss of generality we restrict ourselves to the interval of $n$ 
and $n_{x}$ between $0$ and $1$. 

\begin{figure}[htbp]
\epsfxsize=3.5in
\vspace*{-0.3cm}
\hspace*{-0.5cm}
\epsfbox{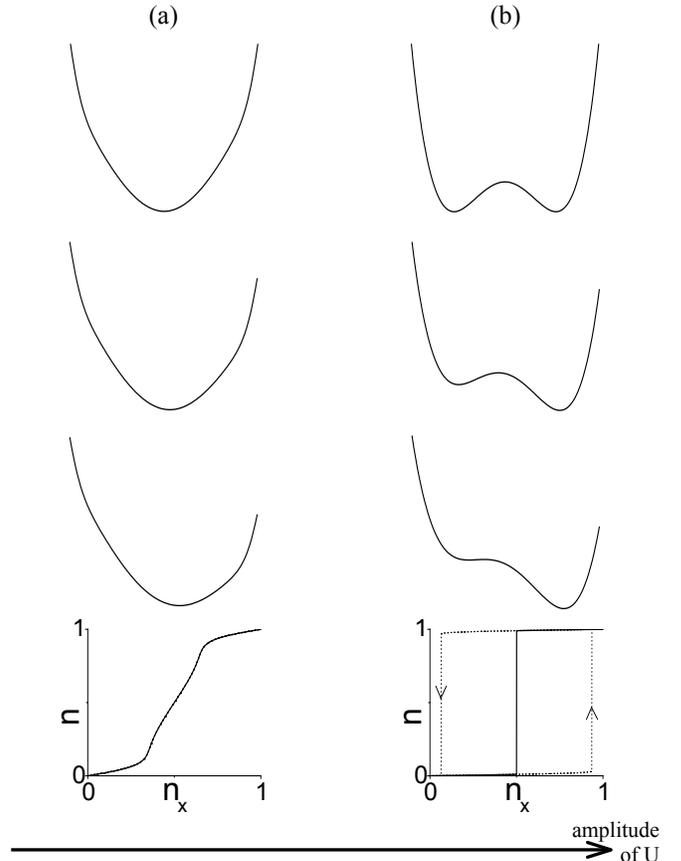}
\vspace*{0.1cm}
\caption{Schematic plots of the ground-state energy as a function of the dot 
occupancy $n$ and corresponding $n(n_{x})$ dependencies for case (i) 
(Ising/Kondo) as discussed in the text.  The first three rows in the figure 
represent successively larger values of $n_x$ beginning with
$n_{x} = 1/2$ in the topmost row.  Equilibrium dependence of the dot 
population $n$ on the gate voltage $n_{x}$ is presented by solid curves in 
the sketches of the bottom row.  The hysteretic parts of the $n(n_{x})$ 
dependencies are shown by dotted lines, and the arrows on the hysteresis loops 
reflect the direction of the change of $n_{x}$.}
\end{figure} 

In order to compute the $n(n_{x})$ dependence we need to minimize the 
full ground-state energy $E(n, n_{x})$ of the system, in principle 
computable from (\ref{Action}) and (\ref{Clenergy}).  Qualitatively
the outcome can be anticipated without calculation by 
noticing that akin to $E_0(\phi, n_{x})$, the ground-state energy 
$E(n,n_{x})$ as a function of $n$ must be a sum of a periodic and parabolic 
functions.  Here there are two main cases to consider.

\vskip .1cm
\noindent {\bf Case (i):}
In the first case the maxima of the periodic part of $E$ are sufficiently 
sharply peaked and the ground-state energy $E$ as function of $n$ in the 
interval $0 \le n \le 1$ sees either one or two minima.  Fig.1 illustrates 
this case.  The top row shows two possible sketches of the 
ground-state energy as a function of $n$ at the degeneracy point 
$n_{x} = 1/2$, and two subsequent rows demonstrate how these curves tilt as 
$n_{x}$ increases.  The solid curves at the bottom row show equilibrium 
$n(n_{x})$ dependencies found by minimizing the ground-state energy.  

\begin{figure}[htbp]
\epsfxsize=3.4in
\vspace*{-0.3cm}
\hspace*{-0.35cm}
\epsfbox{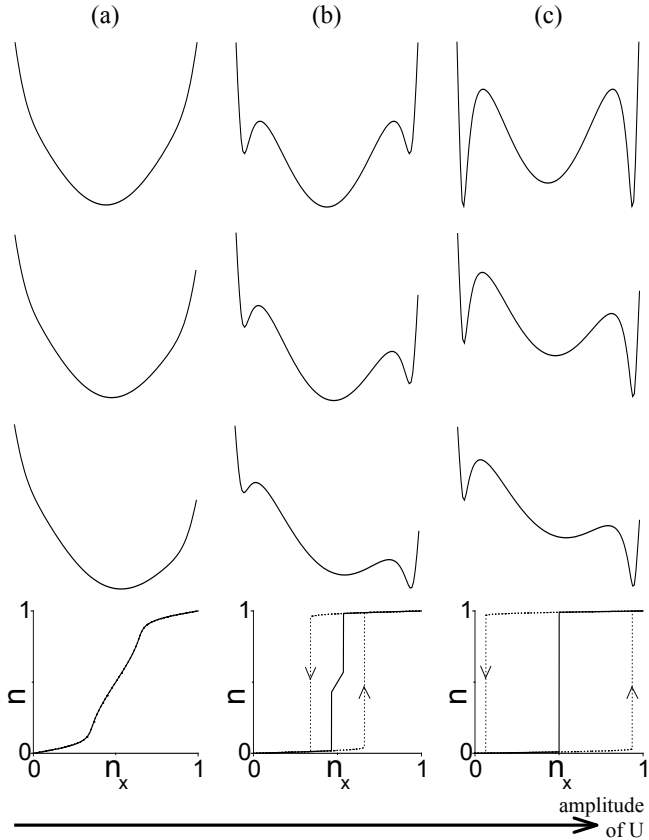}
\vspace*{0.1cm}
\caption{Schematic plots of the ground-state energy and corresponding 
$n(n_{x})$ dependencies in case (ii) (tricritical).  The plot is organized as 
in Figure 1.}
\end{figure} 

In the first scenario, pictured in Fig.1a, there is a single minimum in $E$, 
the position of which smoothly changes as $n_x$ is
varied.  The population of the dot, $n(n_{x})$, is then continuous and  
will look as sketched in the bottommost figure of Fig.1a.  
In the presence of two minima, pictured in Fig.1b, a modified Coulomb 
staircase is recovered upon adiabatic variation of $n_{x}$.  This staircase 
differs from its classical counterpart \cite{Averin&Likharev} in that the 
population plateaus will have a nonzero slope and the jump in dot population 
occuring at $n_x =1/2$ between the two degenerate minima is 
{\it less} than unity.  The inter-plateau jumps are first-order phase 
transitions, and as such they can be accompanied by hysteretic phenomena.  
Indeed if one changes $n_{x}$ non-adiabatically, the system stays at the left 
well until the well ceases to exist and only then will it jump into the 
remaining right well.  The resulting $n(n_{x})$ dependence will deviate from 
the equilibrium staircase.  Upon decreasing $n_{x}$ a different path, 
$n(n_{x})$, will be taken and a hysteresis loop will be produced as shown in 
the bottommost sketch of Fig.1b by the dotted lines with arrows.
   
These two scenarios are distinguished primarily by the amplitude of the 
periodic part, and one can imagine going between the staircase and continuous 
$n(n_{x})$ dependencies by closing or opening the dot.  Alternatively however,
one can imagine varying the reservoir interactions (varying $g$) or  
changing the dot size, i.e. its capacitance $C$.

\vskip .1cm
\noindent {\bf Case (ii)}:  
If the maxima of the periodic part of $E$ are sufficiently flat, 
the ground-state energy, $E$, as a function of $n$ can have as many as 
{\it three} distinct minima.  Fig.2, in a fashion similar to Fig.1,
illustrates this more complicated case.  The three cases presented
in Fig.2 are connected by varying the amplitude of the periodic part of the 
energy.

In the column (a) of Fig.2, the amplitude of the periodic part
is weak and the ground-state energy has a single minimum for any $n_{x}$.  
This results in a continuous dependence $n(n_{x})$ depicted at the bottom of 
Fig. 2a.  As the amplitude of $U$ is increased, one arrives
at column (b) of Fig. 2.  When the gate voltage is tuned to 
degeneracy point, $n_{x} = 1/2$, the ground-state energy has a global minimum
at $n = 1/2$ and degenerate metastable minima at $n$ close to $0$ and $1$.  
If one increases $n_{x}$ adiabatically away from $1/2$, the number of 
electrons on the dot changes continuously (from a half-integer value) until 
the depths of the ``half-integer'' and the right
``integer'' minima become equal.  This happens at a critical value
of $n_{x}$ in the range, $1/2 < n_{x} < 1$, and a first-order phase transition 
onto the ``integer'' plateau occurs.  Upon further increasing $n_{x}$, the 
state of the dot changes continuously until the next ``half-integer'' minimum 
comes into play.  As a result, a staircase with both integer and half-integer 
plateaus is produced, as sketched at the bottom of column (b).  We will arrive 
at qualitatively the same $n(n_{x})$ dependence if ``integer'' minima are 
not present at the top energy curve but develop and become global in tuning 
the gate voltage away from $n_{x} = 1/2$.  

As the amplitude of the periodic part of the potential is increased further 
one arrives at column (c) of Fig. 2.  Here the ground-state energy has 
degenerate minima at $n$ close to $0$ and $1$ and a metastable minimum at 
$n = 1/2$.  If initially the dot state corresponds to the leftmost well, and 
one starts adiabatically increasing $n_{x}$ away from $1/2$ the energy curve 
tilts, and the system jumps into the rightmost well.  Upon further increase 
of $n_{x}$ the average number of electrons on the dot increases continuously 
until the next degeneracy point is reached.  As a result a modified Coulomb
staircase, as given at the bottom of column (c) in Fig. 2, is recovered and the
presence of the third metastable minimum does not qualitatively affect 
the dot occupancy (compare with Fig. 1b).  Hysteresis is also possible in this 
case as shown in Figs.2b and 2c.

We can reconsider both case (i) and (ii) from a unified viewpoint by focusing 
solely on the charge degeneracy points $n_{x} = 1/2$.  
Qualitatively $E_{0}(\phi, 1/2)$ appears as in the top rows of Figs.1 and 2 
and can be described by 
$$
E_{0} \approx - u_{2}s^{2} + u_{4}s^{4} + u_{6}s^{6},
$$
with $s = \phi - 1/2$ (the coefficients $u_{2,4}$ can be of both signs while 
$u_{6} > 0$).  When employed in (\ref{Action}), we arrive 
at a one-dimensional continuous-spin Ising system with $1/\tau^{2}$ 
ferromagnetic exchange interactions.  A small deviation from the charge 
degeneracy points can be modeled by applying a weak fictitious magnetic field 
$h \sim n_{x} - 1/2$, i.e. by adding a $-hs$ term to the expansion 
(an equivalent argument using the language of the Kondo model has been given 
by Matveev \cite{Matveev91,Matveev95}).  
If $u_4 > 0$, the sixth order term can be ignored and we obtain the standard 
long-ranged Ising model, corresponding to case (i) discussed above.  With 
$u_4 < 0$ stability demands we keep $u_6$, and we obtain a long-ranged version 
of the tricritical Ising model \cite{Lubensky&Chaikin} and so reproduce the 
findings of case (ii).

We note that recently, while studying the problem of occupation of a 
resonant level (for example, an impurity or a small quantum dot) coupled to a 
Luttinger liquid, Furusaki and Matveev \cite{FurMat} also noticed a 
connection to the classical one-dimensional Ising model with interactions 
decaying as inverse square of distance. 

In the next two sections we consider the cases (i) and (ii) in more 
quantitative detail.

\section{KONDO/ISING TYPE}

In this section we consider a dot whose potential, $E_{0}(\phi )$, 
can be approximated by a two-well potential and is thus in the ordinary 
(as opposed to tricritical) long-ranged Ising universality class.  The 
truncation of charge fluctuations of the quantum dot to that of a two-level 
system performed in \cite{Matveev91} maps the dot problem to the conventional 
Ising model with long-ranged $1/\tau^{2}$ interactions \cite{Thouless} or 
equivalently the single-channel Kondo model \cite{Anderson&Yuval}.

The physical consequences of this equivalence can be derived from an 
examination of the renormalization-group flows of the long-range
Ising/Kondo model.  To understand these flows we follow the treatment of the
dual dissipative SQUID problem \cite{Bray&Moore}, and introduce two parameters,
the distance between the two wells of the potential $q$, and the 
(dimensionless) inter-well rate of tunneling $\Delta$.  Under a 
renormalization-group transformation these parameters flow 
at lowest order as follows \cite{Anderson&Yuval,Bray&Moore}:
\begin{equation}
\label{two-state}
{d\Delta\over d\ln(\Lambda t)} = (1 - q^{2}/g)\Delta,
\hspace{1.0cm} 
{dq^{2}\over d\ln(\Lambda t)} = - q^{2}\Delta^{2} .
\end{equation}
Here $t$ is the running scale and the equations describe how the two 
parameters renormalize upon integrating out high-frequency modes.  
The initial conditions, $\Delta_{0}$ and $q_{0}$, can be computed 
from the form of the action in (\ref{Action}).  An example of such a
computation done in the SQUID context may be found in \cite{Bray&Moore}.

The flow diagram corresponding to (\ref{two-state}) is sketched in Fig. 3.
\begin{figure}[htbp]
\epsfxsize=3.6in
\vspace*{-0.3cm}
\hspace*{-0.7cm}
\epsfbox{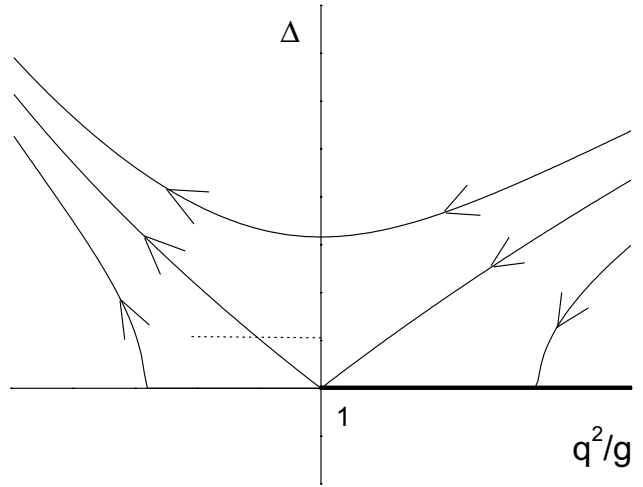}
\vspace*{0.1cm}
\caption{The flow diagram corresponding to (\ref{two-state}).  The stable 
part of the $\Delta = 0$ fixed line is shown 
in bold, and the arrows indicate the direction of the flow.  The locus of the
free-electron models, $g = 1$, is presented schematically by the dotted 
line.}  
\end{figure} 
\noindent
The flows can be divided into two main regions, corresponding to
the ordered and the disordered phases of the Ising model.

In the former case are
systems satisfying $q_{0}^{2}/g \ge 1$ and 
$\Delta_{0}^{2}/2 \le q_{0}^{2}/g - 1 - \ln(q_{0}^{2}/g)$, i.e. those to the
right of the separatrix  $\Delta^{2}/2 = q^{2}/g - 1 - \ln(q^{2}/g)$, are 
carried by the flow to a stable fixed line, $\Delta = 0$, $q^{2}/g \ge 1$.
The absence of tunneling between the two minima of the potential
indicates that we are in the ordered phase of the Ising model.  
The functional form of $n(n_{x})$ is as shown in Fig.1b, and the dot 
population discontinuity, $\Delta n$ (given by $q$), at $n_{x} = 1/2$ 
satisfies the inequality
\begin{equation}
\label{inequality}
\Delta n \ge \ g^{1/2}.
\end{equation}
The equality is reached at the Ising phase transition 
point, $\Delta_{0}^{2}/2 = q_{0}^{2}/g - 1 - \ln(q_{0}^{2}/g)$, where the
population jump $\Delta n = g^{1/2}$ depends {\it only} on the properties of 
the reservoir.  This universal relationship is analogous to that between the 
magnetization jump, the amplitude of inverse-square interaction and the phase 
transition temperature of the Ising model 
\cite{Thouless,Anderson&Yuval,Bray&Moore,LM}.     

Although systems to the left of the separatrix 
$\Delta^{2}/2 = q^{2}/g - 1 - \ln(q^{2}/g)$ are taken by the flow outside of
the range of applicability of Eqs. (\ref{two-state}), the effective
tunneling, $\Delta$, between the two minima of the two-well potential
can be seen to increase.
We are thus in the disordered phase of the Ising model.  Here the 
dependence, $n(n_{x})$, is continuous as in Fig.1a.  Although the Ising 
transition is continuous, in going from Fig.1a to Fig.1b, the $n(n_{x})$ 
dependence changes {\it discontinuously}.  In the disordered phase, the 
ground-state energy slightly away from $n_{x} = 1/2$ is 
given by 
$$E = (\pi \hbar/ g\xi)<s>^{2} - h<s> ,
$$ 
where the dependence on the correlation length $\xi$ is dictated by the 
one-dimensional nature of the problem.  Minimization with respect to $<s>$ 
leads us to the equation,
\begin{equation}
\label{step}
\langle s \rangle \equiv n - 1/2 \sim (g\xi/(2\pi\hbar ))(n_{x} - 1/2),
\end{equation}
describing the shape of the smeared population step.

In classical limit of closed dot, $\Delta_{0} = 0$, the distance between 
the minima of the two-well potential is unity, $q_{0} = 1$, which implies 
that if the electrons in the constriction are repulsive, $g < 1$, these 
systems belong to the ordered phase of Ising model.  Allowing weak tunneling, 
i.e. sufficiently small $\Delta_{0}$, renormalizes the population step 
downwards, and we arrive at the $n(n_{x})$ dependence of the kind shown in 
Fig.1b.  At sufficiently large critical $\Delta_{0c}$ the Ising phase 
transition takes place, and for $\Delta_{0} > \Delta_{0c}$ the dependence 
$n(n_{x})$ is continuous, as in Fig.1a.  

In this same classical limit, $\Delta_{0} = 0$, 
but with the electrons in the 
constriction attractive, 
$g > 1$, we find ourselves on the unstable part of the $\Delta = 0$ fixed 
line.  Here allowing infinitesimally small tunneling, $\Delta_{0}$, 
immediately destroys the classical staircase, thus leading to
the continuous $n(n_{x})$ dependence, Fig.1a.  For $g > 1$, the flow 
starts out vertically away from the $\Delta = 0$ line.  In order to compute 
$\xi$ to lowest order in $\Delta_{0}$, the renormalization of $q$ given by 
the second of Eqs.(\ref{two-state}) can be ignored.  Integrating the first of 
Eqs.(\ref{two-state}) up to a scale at which $\Delta \approx 1$ determines 
the correlation length to be 
$\xi \approx \Lambda^{-1} \Delta_{0}^{g/(1-g)}$.  Physically the difference 
between the repulsive and attractive cases can be understood by noticing that 
interparticle attraction enhances quantum fluctuations in the constriction,  
thus making it easier for the electrons to enter and leave the dot; the effect
of repulsion is opposite. 
    
The free-electron case, $g = 1$ is the most delicate to treat because 
it separates regimes which are stable and unstable against perturbations of
infinitesimally weak tunneling.  If the function $U(\phi)$ is a periodic comb 
of delta-function wells of large but finite amplitude, then one still has 
$q_{0} = 1$.  Solving Eqs.(\ref{two-state}) with the initial condition 
$q_{0} = 1$ and $\Delta_{0} \ll 1$, we find 
$\xi \approx \Lambda^{-1}\exp(\pi/2\Delta_{0})$.  If this $\xi$ is inserted in 
(\ref{step}), then the shape of the smeared population step to leading 
order in $\Delta_{0}$ coincides with that given in Ref. \cite{Matveev91}.
For all other periodic functions $U(\phi)$ one has $q_{0} < 1$, and the locus 
of the corresponding models is shown in
Fig.3 by the dotted line.  If the initial values $q_{0}$ and $\Delta_{0}$ are
not below the outgoing, leftward separatrix in Fig. 3, then 
$\xi \approx \Lambda^{-1}\exp(c_{1}/\Delta_{0})$ where the constant $c_{1}$ 
interpolates between $\pi/2$ (the $q = 1$ axis) and unity (the left 
separatrix).  If the initial values are below the separatrix, then one has 
$\xi \approx \Lambda^{-1}\exp[(c_{2}/\Delta_{0})\ln\Delta_{0}^{-1}]$ where 
$c_{2}$ is another constant.  To conclude, if the electrons in the 
constriction are noninteracting {\it and we are in the Kondo/Ising universality
subclass}, then the $n(n_{x})$ dependence is {\it always} continuous, as 
shown in Fig.1a, for arbitrarily small tunneling coupling \cite{Matveev91}.  

\section{TRICRITICAL ISING TYPE}  

The results derived so far examine the case where at $n_{x} = 1/2$ the 
effective potential energy can be approximated by a two-well 
potential.  But this assumption is invalid for potentials whose periodic 
part has sufficiently flat maxima and thus whose energy dependencies
appear qualitatively similar to the top row of Fig.2.  Moreover, if even the 
two-state assumption is correct for the bare potential, renormalization of 
truncated two-state potential by zero-point motion might be qualitatively 
different from renormalization of the full potential.  In such cases the 
tricritical behavior as pictured in Fig.2 will be overlooked.

Therefore it is important to reexamine the problem without making the 
two-state assumption.  In this section we consider exactly such a case where 
approximating the potential energy as two wells is inappropriate.
In what follows we choose the periodic part of the potential 
entering (\ref{Clenergy}) to be $U(\phi) \sim \cos2\pi\phi$, and rewrite
(\ref{Clenergy}) as
\begin{equation}
\label{sgclenergy}
E_{0}(\phi, n_{x}) = {\Lambda\hbar\over g}[\pi b (\phi - n_{x})^{2} - 
{v\over2\pi}\cos2\pi\phi].
\end{equation}
Here $b = e^{2}g/(2\pi C\Lambda \hbar)$ is the dimensionless charging energy, 
while the dimensionless parameter $v$ in the limit $v \ll 1$ can be expressed 
in terms of the reflection amplitude at the point contact\cite{Matveev95}.  
This potential (\ref{sgclenergy}) models Coulomb blockade \cite{Matveev95}, 
SQUIDs \cite{Likharev}, and multilayer adsorption phenomena
\cite{Weeks,Huse,Weichman&Prasad}.  Surprisingly we find 
that the model (\ref{sgclenergy}) {\it does not} belong to the 
Kondo/Ising universality subclass although na\"{\i}vely
for $n_{x} = 1/2$ the two-state assumption would appear to be reasonable.

This model is readily understood if the classical limit, $g \rightarrow 0$, is
taken in the constriction. Then the fluctuations of the field, $\phi$, 
controlled by the first term of (\ref{Action}) are frozen.  To find the 
$n(n_{x})$ dependence, we have to minimize (\ref{sgclenergy}) with respect 
to $\phi = n$.  The outcome is known from the SQUID context \cite{Likharev}: 
for $v > b$ the $n(n_{x})$ dependence qualitatively looks like that shown in 
Fig. 1b (hysteretic SQUID regime).  If, on the other hand, $v < b$, 
the $n(n_{x})$ dependence is of the kind presented in Fig. 1a.  The phase 
transition at $v = b$ is continuous with the population jump which grows from 
zero as one enters the staircase phase.

\subsection{Scaling analysis}

To understand the role of quantum fluctuations in the constriction 
(nonzero $g$) we first treat $E_{0}(\phi,n_{x})$
perturbatively (i.e. $v$ small, putting us in the high-transparency limit).  
The corresponding lowest-order renormalization-group equations
\begin{equation}
\label{pertrg}
{d{\tt v}\over d\ln(\Lambda t)} = (1 - g){\tt v} ,
\hspace{1.0cm} 
{d{\tt b}\over d\ln(\Lambda t)} =  {\tt b} ,
\end{equation}               
parallel those given by Huse \cite{Huse} in the context of the adsorption 
phenomena.  The parameters $v$ and $b$ of (\ref{sgclenergy}) serve as initial 
values to (\ref{pertrg}).  We note that due to nonanalytic $|\omega|$ 
dependence in (\ref{Action}) there is no renormalization of $g$ to any order.
This is in contrast to the adsorption phenomena where the corresponding 
renormalization of an interface stiffness coefficient is an important part of 
physics.  

With $b=0$, the 
first equation in (\ref{pertrg}) is known to describe the 
physics of a Luttinger liquid in the
presence of a weak point inhomogeneity \cite{Kane&Fisher}: there is a phase
transition at $g = 1$ which separates insulating, $g < 1$, from  
conducting regimes, $g > 1$.  The second equation, solely due to the 
scaling transformation, tells us that the charging energy is always a relevant 
perturbation.  In $g \le 1$ regime, $v$ does not decrease under the flow, 
and Coulomb blockade can potentially occur.
In this regime integrating the first equation of (\ref{pertrg}) 
until $ v \approx 1$ provides us with a length scale 
$\xi_{v} \approx \Lambda^{-1} v^{1/(g - 1)}$, i.e. a 
longitudinal correlation length \cite{KS}.  In the
absence of the charging term, at scales exceeding $\xi_{v}$, the insulating 
behavior sets in and the fluctuations of the field, $\phi$, are suppressed. 

The presence of the charging energy introduces another length scale into the 
problem, $\xi_{b} \approx (\Lambda b)^{-1}$, as can be inferred either from 
(\ref{Action}) and (\ref{sgclenergy}), or from integrating the second equation
of (\ref{pertrg}).  In the absence of the periodic term at the scales 
exceeding $\xi_{b}$, the fluctuations of $\phi$ are suppressed.

The evolution of the Coulomb blockade as a function of the system parameters 
can be understood qualitatively as a result of the interplay between the 
scales $\xi_{v}$ and $\xi_{b}$ (both assumed to be much larger than the
microscopic scale $\Lambda^{-1}$).  If  $\xi_{b} \lesssim \xi_{v}$ the 
fluctuations of $\phi$ are suppressed by the
charging term before the periodic term comes into play.  This implies that 
the physics is dominated by the charging term, the dot population $n$ tracks
the gate voltage $n_{x}$ with electronic discreteness a small effect.  The 
$n(n_{x})$ dependence is continuous as in Figs.1a and 2a.  If, on the other 
hand, $\xi_{b} \gtrsim \xi_{v}$, the fluctuations are suppressed by the 
periodic term before the charging effects come into play.  This means that 
the physics is dominated by the particle discreteness with $n(n_{x})$ 
correspondingly in the form of a staircase.  The two regimes are separated by 
a phase transition occuring at $\xi_{b} \approx \xi_{v}$, i.e. for
\begin{equation}
\label{endpoint}
vb^{g - 1} \approx 1 
\hspace{0.5cm}
{\rm or}
\hspace{0.5cm}
g \approx 1 - {\ln(1/v)\over \ln(1/b)}.
\end{equation}  
We note that (\ref{endpoint}) has roughly the correct limit $b \approx v$ as 
$g \rightarrow 0$, and implies that for $g = 1$ a phase transition occurs at 
some $v$ which is $b$-independent in the limit $b \rightarrow 0$. 
We now show that the scaling condition in (\ref{endpoint}) underspecifies the 
physics and that there are in fact {\it two} phase transitions occuring when 
$\xi_b \sim \xi_v$.

\subsection{Variational analysis}

For a more quantitative analysis we use Feynman's \cite{Feynman} variational 
principle for the ground-state energy:
\begin{equation}
\label{var}
E \le E_{1} = {\tt E}_{0} + (T/\hbar)<A - A_{0}>_{0}
\end{equation}
where $T$ is the temperature, and $\hbar/T$ has a meaning of the system size
in the $\tau$ direction; the $T = 0$ limit will be taken at the end.  The 
notation $<>_{0}$ denotes expectation values 
computed using an arbitrary reference
action $A_{0}$, and ${\tt E}_{0}$ is the ground-state energy corresponding to
$A_{0}$.  

This method has been remarkably successful in analyzing the 
roughening phase transition \cite{Saito}, multilayer adsorption phenomena 
\cite{Weeks}, wetting transitions \cite{LKZ}, and a variety of problems of 
quantum mechanics and quantum-field theory \cite{Kleinert}.  
Perhaps most relevant to our purposes, has been the application
of the variation principle to the problem
of quantum Brownian motion in a periodic 
potential \cite{Fisher&Zwerger} which is mathematically identical to the 
single-impurity problem \cite{Kane&Fisher} and formally the $b = 0$ limit of 
the Coulomb blockade problem (\ref{Action}), (\ref{Clenergy}).  Here the 
variational method provides exact description of the phase transition at 
$g = 1$ \cite{KS}. 

It is physically reasonable to select the trial action $A_{0}$ in a 
Gaussian form similar to that in \cite{Weeks} and \cite{Fisher&Zwerger}:
\begin{equation}
\label{trialaction}
A_{0} = {\pi\ \hbar \over g} 
[\int \limits _{- \Lambda}^{\Lambda}{d\omega \over 2\pi}|\omega|
|\phi(\omega)|^{2} + \Lambda m \int d\tau (\phi - n)^{2}].
\end{equation}
The dimensionless variational parameters which include the familiar number of
electrons on the dot $n$ and a new parameter $m$ (controlling the extent of
fluctuations about $n$) are selected to minimize $E_{1}$ in (\ref{var}). 

Using (\ref{trialaction}) the upper bound $E_{1}$ entering (\ref{var}) can be 
computed.  Ignoring unimportant additive and overall factors we have
\begin{eqnarray}
\label{varenergy}
E_{1}& \sim& (1 + b)\ln(1 + m) - b\ln(m)\nonumber\\ 
&+& g^{-1}[2\pi^{2}b(n - n_{x})^{2} - 
v({m\over 1 + m})^{g}\cos2\pi n] .
\end{eqnarray}
Minimizing $E_{1}$ with respect to $m$ and $n$ we arrive at the system of two
equations
\begin{equation}
\label{meq}
m - b = v({m\over 1 + m})^{g}\cos2\pi n
\end{equation} 
\begin{equation}
\label{neq}
2\pi b(n - n_{x}) + v({m\over 1 + m})^{g}\sin2\pi n = 0
\end{equation}
which are mathematically identical to those of Weeks \cite{Weeks} derived in 
the context of adsorption phenomena.  

The most valuable feature of the variational approach is its nonperturbative 
nature.  The perturbative renormalization-group treatment (\ref{pertrg}) is 
included as a special case as can be seen from an argument which parallels 
that of Br{\'e}zin, Halperin, and Leibler originally given in the context of 
wetting transitions \cite{BHL}:

Let us split the fluctuating number field $\phi$ into a ``slow'', $\phi_{0}$, 
and a ``fast'' part, $\phi_{1}$, so that $\phi = \phi_{0} + \phi_{1}$, and 
``integrate out'' fast degrees of freedom $\phi_{1}$ treating 
(\ref{sgclenergy}) as a perturbation.  This turns (\ref{sgclenergy}) into its 
renormalized (but unrescaled) counterpart
\begin{equation}
\label{renenergy}
E_{L} = {\Lambda\hbar\over g}[\pi b (\phi - n_{x})^{2} - 
{v(L\Lambda)^{-g}\over2\pi}\cos2\pi\phi] ,
\end{equation}
where the subscript $0$ is dropped for brevity, and at the scale 
$L$ all high-frequency fluctuations are integrated out.  We note that equations
(\ref{pertrg}) and (\ref{renenergy}) are equivalent in terms of the physics
they describe.  The number of electrons 
$n$ on the dot, and the correlation length $L$ can be found from the conditions
similar to those of \cite{BHL},
\begin{equation}
\label{bhl}
[\partial E_{L}/\partial\phi]_{n} = 0,
\hspace{0.3cm} 
(2\pi \Lambda \hbar /g)(L\Lambda)^{-1} 
= [\partial^{2} E_{L}/\partial\phi^{2}]_{n},
\end{equation}     
where the dependence on $L$ in the second equation is dictated by the 
$|\omega|$ dependence in (\ref{Action}).  It is straightforward to verify that 
substituting (\ref{renenergy}) in (\ref{bhl}) produces two equations which are
just the $m \ll 1$ limit of (\ref{meq}) and (\ref{neq}) provided the 
identification $m = (L\Lambda)^{-1}$ is made. 

To summarize, the $m \ll 1$ limit of the variational analysis is equivalent to
the perturbative renormalization-group treatment.  

Below we analyze the coupled system of equations (\ref{meq}) and (\ref{neq}),
and whenever multiple solutions are found, the one giving the lowest 
ground-state energy (\ref{varenergy}) is selected. 
We make two preliminary observations.
Firstly we note that solving (\ref{meq}) provides us with a periodic 
$m(n + 1) = m(n)$ dependence which after substituting into (\ref{varenergy}) 
results in a function which is a sum of a periodic and parabolic functions
as we already anticipated.  
Secondly, for $n_{x}$ 
integer or half-integer Eq.(\ref{neq}) has a solution 
$n = n_{x}$ as expected.
 
\subsection{Free electrons in the constriction:  ${\bf g = 1}$}

For $g = 1$, Eq. (\ref{meq}) admits an analytic solution,
and thus the ground-state energy (\ref{varenergy}) can be explicitly 
computed.  The results are simplest in the limit $m \ll 1$.  
In this limit the solution to (\ref{meq}) is given by
\begin{equation}
\label{msol}
m = {b \over 1 - v\cos2\pi n},
\end{equation}
valid for $b \ll 1$ and $v < 1$.  Equation (\ref{msol}) shows that
fluctuations around integer dot population are always smaller compared to
those for a half-integer number of electrons on the dot.  Substituting
(\ref{msol}) into (\ref{neq}) we find the $n_{x}(n)$ dependence 
which parallels the result of Weeks \cite{Weeks}:
\begin{equation}
\label{nsol}
n_{x} = n + {v\sin2\pi n \over 2\pi(1 - v\cos2\pi n)}.
\end{equation}       
The remarkable feature of (\ref{nsol}) is its {\it universality} - there is 
no dependence on the dot capacitance in the $b \ll 1$ limit.  We emphasize 
that the result (\ref{nsol}) goes beyond the original model 
(\ref{sgclenergy}):  any even periodic function $U(\phi)$ from 
(\ref{Clenergy}) can be decomposed into a Fourier series consisting of cosine 
functions.  For $g = 1$ the leading harmonic [accounted for by
(\ref{sgclenergy})] responsible for (\ref{nsol}) is marginal [see the first of
Eqs.(\ref{pertrg})] while higher harmonics are irrelevant.  In the limit 
$b \rightarrow 0$ the contribution of higher-order harmonics vanishes 
(unless the corresponding Fourier coefficients are unreasonably
large which we assume to be not the case).     

At $g=1$ the variational argument gives two critical values of the
dimensionless amplitude, $v_{c1}$ and $v_{c2}$: the lower 
$v_{c1}$
marks the onset of the regime with half-integer plateaus while the upper 
$v_{c2}$
gives the transition to a regime solely with integer plateaus, i.e. 
a modified Coulomb staircase.  

To determine $v_{c1}$ one can study the monotonicity
of the $n_x(n)$ function as given in (\ref{nsol}).
For sufficiently small $v$, the $n_x(n)$ dependence is monotonic.  For 
$v \ll 1$ solving (\ref{nsol}) to lowest order in $v$ reproduces the result of 
Matveev
\cite{Matveev95}, 
$$
n = n_{x} - (v/2\pi)sin2\pi n_{x}.
$$  
As $v$ becomes larger, $n_{x}(n)$ ceases to be monotonic.  Defined when
$dn_{x}/dn$ vanishes, this happens at
$v_{c1} = \sqrt{3}/2 < 1$ and $\cos2\pi n_{c} = 1/\sqrt{3}$ \cite{Weeks}. 
The corresponding gate voltage $n_{x}$ is {\it not} a half-integer, 
a first indication that the lower critical value $v_{c1}$ separates continuous,
Fig. 2a, from discontinuous behaviour, Fig. 2b, with population plateaus 
centered both around the integer and half-integer number of electrons on the 
dot.      

To better understand the phase diagram we look at the ground-state energy.  
Combining (\ref{varenergy}) and (\ref{meq}) for $g = 1$, with (\ref{msol}), 
taking the limit $m, b \ll 1$ and dropping unimportant additive constants we 
arrive at the form,
\begin{equation}
\label{varenergyg=1}
E_{1}(n,n_{x}) \sim b[\ln(1 - v\cos2\pi n) + 2\pi^{2}(n - n_{x})^2].
\end{equation}
Comparing the periodic parts of (\ref{varenergyg=1}) and (\ref{sgclenergy}) we
conclude that quantum fluctuations {\it sharpen} integer minima and 
{\it flatten} half-integer maxima.
The minimization of $E_{1}$ with respect to $n$ reproduces (\ref{nsol}).  
For $v < v_{c1}$ the 
ground-state energy $E_{1}$ always has a unique minimum which results in a
continuous $n(n_{x})$.  For $v$ just slightly above $v_{c1} = \sqrt{3}/2$ 
and $n_{x} = 1/2$ the ground-state energy still has a global minimum 
at half-integer dot population.  But upon deviating from  
$n_{x} = 1/2$, a near integer minimum develops and eventually becomes global. 
The phase transition at $v_{c1}$ is continuous. 

As $v$ is increased beyond $\sqrt{3}/2$, we come to the second
transition.  
The approach to this transition is characterized by 
the deepening of two metastable minima close to $n = 0$ and $n = 1$ 
(Fig.2b) with $n_{x} = 1/2$.
The critical value $v_{c2}$ marks the point at which two near integer minima 
become equal in depth to the half-integer minimum.  This is a first-order 
transition between the regimes sketched in Figs.2b and 2c.  The second 
critical value can be estimated accurately analytically from the condition
$$
E_{1}(1/2, 1/2) = E_{1}(0, 1/2),
$$ 
resulting in 
$v_{c2} \approx \tanh(\pi^{2}/4) \approx 0.9857 < 1$.  The true $v_{c2}$ is 
marginally smaller as the competing 
minimum is shifted slightly away from $n = 0$.  For 
$v_{c2} < v < 1$ there are only population plateaus centered around integer
number of electrons on the dot, although the ground-state energy always has
a meta-stable minimum corresponding to a half-integer dot population.  

Examples of the three regimes separated by $v_{c1}$ and $v_{c2}$ are plotted in
Fig.2 with values of $v=0.8,0.97$, and $0.988$ for (a)-(c) respectively.
The continuous segments of the $n(n_{x})$ plots were found by inverting 
(\ref{nsol}) while the location of the first-order jumps was determined from 
the minimization of (\ref{varenergyg=1}).  The hysteretic behavior in 
Figs.2b and 2c arises when the inversion leads to multiple solutions and was 
determined by continuing the monotonic regime solution beyond the range where
it is energetically favorable.  

It is possible to determine the slopes $dn/dn_{x}$ of the plateaus (given
that that the slope is non-zero, the language of plateaus is used
figuratively).  The slope of the integer plateaus given by $1 - v$ is rather 
flat in the range $v_{c1} = \sqrt{3}/2 < v < 1$.   
This flatness is a direct consequence of the sharpness of the near integer 
minima of the energy curves.  
For the half-integer plateaus over this same range of $v$ we find the slope 
to be much larger, $1+v$.  This steepness is a consequence of the shallowness 
of the half-integer minima in the energy curves.  Alternatively one can recast 
this in the language of the tricritical Ising model \cite{Lubensky&Chaikin}.
Two adjacent integer plateaus are then regarded as two ordered Ising phases 
of opposite magnetization.  The half-integer state of the dot
is identified with the disordered (zero-magnetization) phase of 
the Ising model.  On the $n(n_{x})$ curves of Fig.2b the disordered 
phase coexists with both ordered phases.  Equivalently upon increasing $v$, 
the disordered phase disappears gradually in the interval between $v_{c1}$ 
and $v_{c2}$.  On the other hand, for the systems of
the Kondo/Ising type only ordered phases can coexist on the $n(n_{x})$
curves.  Correspondingly, increasing the amplitude of $U$ in (\ref{Clenergy}) 
beyond some critical value leads the disordered phase to disappear altogether.

As was demonstrated in Section IIIB, the $m \ll 1$ limit of the variational 
treatment is equivalent to the perturbative renormalization-group analysis of 
the problem based on Eqs.(\ref{pertrg}) or equivalently 
Eqs.(\ref{renenergy}) and (\ref{bhl}). We speculate on the grounds of the 
following plausibility argument that for the special case of free electrons, 
$g = 1$, such an analysis becomes exact in the limit $b \rightarrow 0$:  

For $b = 0$ the right-hand side of the first of Eqs.(\ref{pertrg}) is 
believed to vanish to {\it all} orders in $v$ {\it if and only if} $g = 1$ 
\cite{Fisher&Zwerger}.  Equivalently this means that on taking the limit 
$g = 1$ and $b = 0$ in (\ref{renenergy}) one would arrive at an exact 
renormalized potential.  On the other hand, in deriving 
Eqs.(\ref{msol}), (\ref{nsol}), 
and (\ref{varenergyg=1}) a small finite $b$ was kept, and the 
$b \rightarrow 0$ limit was taken in the final formulas.  It appears 
reasonable that higher order terms in $b$ not included in (\ref{renenergy}) 
will not affect the results in the $b \rightarrow 0$ limit.   Specifically, 
we argue that Eqs.(\ref{msol}), (\ref{nsol}), and (\ref{varenergyg=1}) solve 
the Coulomb blockade problem exactly for $v < 1$ and $b \rightarrow 0$. 

For $v \ge 1$ the assumption $m \ll 1$ cannot be justified, and we need to 
return to the original system of equations (\ref{varenergy})-(\ref{neq}).
From numerical studies we find that for $b \ll 1$ and $v \ge 1$, the 
$n(n_{x})$ dependence rapidly evolves into a near perfect staircase.

The foregoing analysis presumes $b \ll 1$. 
As $b$ increases, the slope of the integer plateaus, the critical values 
$v_{c1}$ and $v_{c2}$, and the range of existence of the 
intermediate phase with half-integer plateaus all become larger.  
However no qualitative 
changes occur from what was found in the $b \rightarrow 0$ limit.   

\subsection{Interacting electrons in the constriction:  ${\bf g \neq 1}$}

In general we find that our conclusions at $g=1$ are quantitatively
but not qualitatively modified
by the presence of interactions.
Interactions change the level of quantum fluctuations in the system,
making it easier ($g > 1$) or more difficult ($g < 1$) 
for the electrons to enter or leave the dot.  

In the limit $b, vb^{g-1} \ll 1$,  Eqs. (\ref{meq}) and (\ref{neq}) 
give the continuous dependence
\begin{equation}
\label{nsolgneq1}
n = n_{x} - (vb^{g-1}/2 \pi)\sin2\pi n_{x},
\end{equation}
thus generalizing the $g = 1$ result of Matveev \cite{Matveev95}.
If the electrons inside the constriction are attractive, $g > 1$, 
Coulomb blockade effects are substantially weakened.  The range of existence
of the continuous $n(n_{x})$ dependence grows compared to the noninteracting
case; the critical values $v_{c1}$ and $v_{c2}$ grow similarly.  As $g$ goes 
to infinity the number of electrons on the dot $n$ rapidly approaches the gate 
voltage $n_{x}$.
If, on the other hand, the electrons are repulsive, $g < 1$, the effects of 
Coulomb blockade are enhanced.  The range of existence of the continuous 
$n(n_{x})$ dependence shrinks, and the critical values $v_{c1}$ and 
$v_{c2}$ decrease approaching their common $v = b$ limit as $g \rightarrow 0$.

\begin{figure}[htbp]
\epsfxsize=3.8in
\vspace*{-0.3cm}
\hspace*{-1.2cm}
\epsfbox{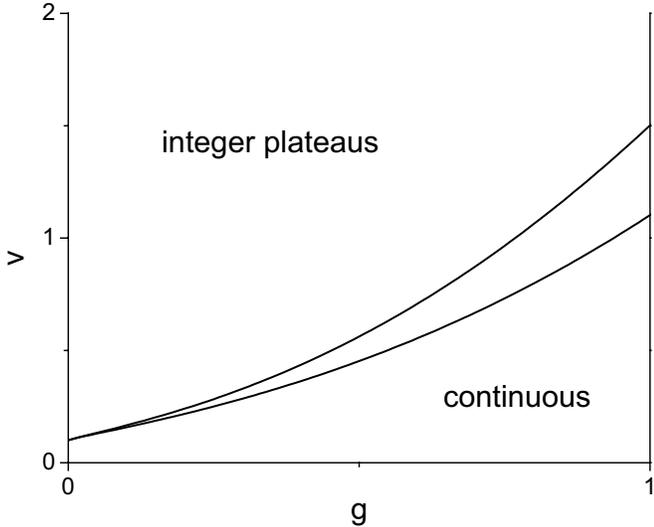}
\vspace*{0.1cm}
\caption{The phase diagram of the model (\ref{sgclenergy}) for $b = 0.1$ and 
$0 \le g \le 1$.  The phase with integer-half-integer sequence of the 
population plateaus exists between the lower, $v_{c1}(g)$, and upper, 
$v_{c2}(g)$, critical curves.}  
\end{figure}

Fig.4  shows the numerically derived phase diagram of the model 
(\ref{sgclenergy}) for $b = 0.1$ and $0 \le g \le 1$.  In between the 
two curves for $v_{c1}$ and $v_{c2}$ lies the phase with half-integer
plateaus.  These two curves meet at $v_{c1}=v_{c2} = b$ in the
limit $g \rightarrow 0$ as expected from (\ref{endpoint}).
In further agreement with the scaling result (\ref{endpoint}) in 
the $b \ll 1$ limit, the critical curves take the form 
$$
v_{c1,2}(g) = a_{1,2}(g)b^{1-g}
$$ 
with  $a_{1,2}(0) = 1$, $a_{1}(1) = \sqrt{3}/2$, and $a_{2}(1) \approx 0.98$.  

We also verified that except for $g = 0$ the $n_{x} = 1/2$ ground-state 
energy (\ref{varenergy}) always has a minimum at half-integer dot 
population:  $g = 0$, $v = b$ is a tricritical point of the model 
(\ref{sgclenergy}). 

Fig.5 shows the numerically derived $n(n_{x})$ dependencies for $g = 1/2$,
$b = 0.01$, and (a) $v=0.13$, (b) $v=0.16$, and (c) $v = 0.2$.  The values 
of $v$ are selected to show continuous (a), and discontinuous $n(n_{x})$ 
dependencies with integer-half-integer (b) and integer-integer (c) sequences of
population plateaus.      
\begin{figure}[htbp]
\epsfxsize=3.5in
\vspace*{-0.3cm}
\hspace*{-0.8cm}
\epsfbox{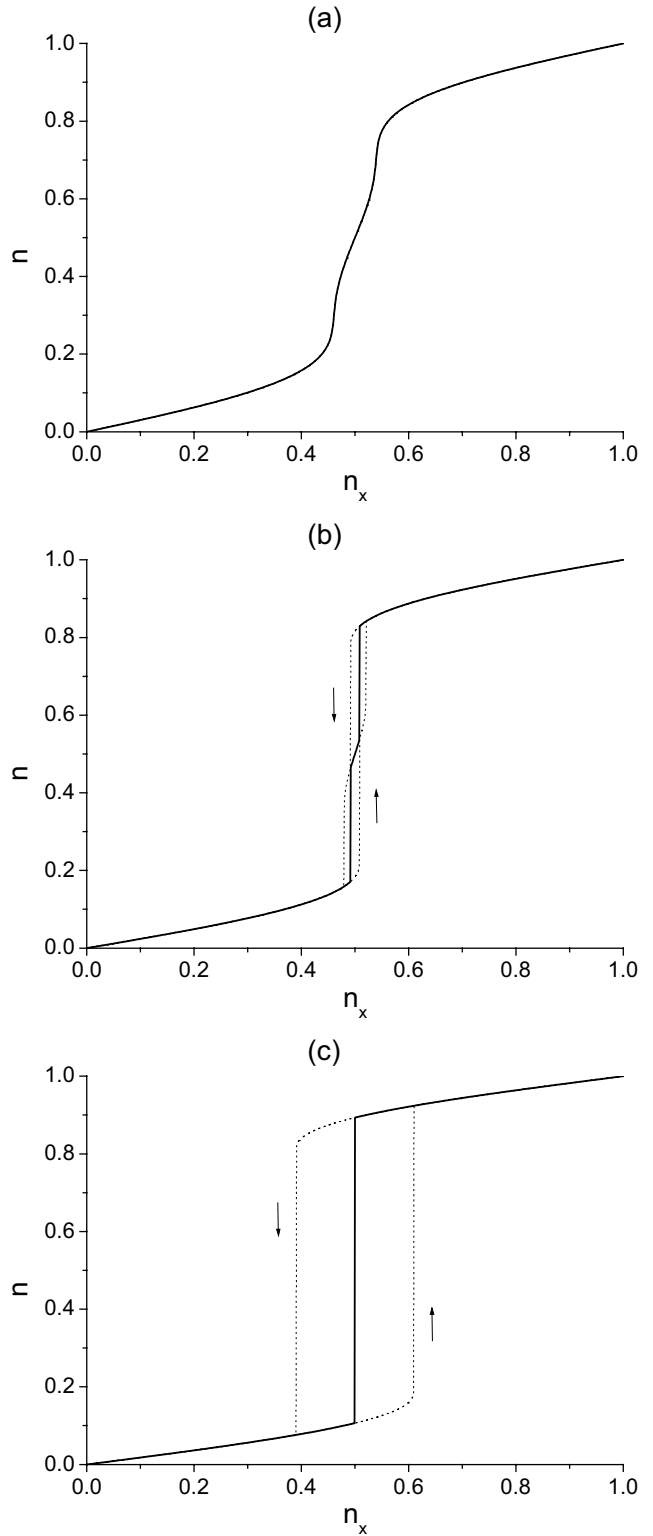}
\vspace*{0.1cm}
\caption{The dependence of electron number $n$ on the dot on the gate voltage
$n_{x}$ for $g = 1/2$, $b = 0.01$, and (a) $v=0.13$, (b) $v=0.16$, and 
(c) $v = 0.2$.  Hysteresis loops are shown by the dotted lines and the 
direction of change of $n_{x}$ is indicated by the arrows.}  
\end{figure}

\section{DISCUSSION}  

The most intriguing prediction of Eqs.(\ref{varenergy})-(\ref{neq}) is the 
existence of a range of parameters where the population of the dot jumps from 
a near integer value to a region of stable population centered about a 
half-integer value.  The origin of this effect is the persistence of the 
half-integer minimum in the population dependence of the ground-state energy 
at the charge degeneracy points.  The effect is quantum - the bare energy
(\ref{sgclenergy}) does not have this property \cite{SilZim}.  
For the effect to occur, the periodic part of the ground-state energy must 
have sufficiently flat maxima which the bare energy (\ref{sgclenergy}) does 
not have.  

The effect is not entirely unexpected and resembles the classical effect 
of an inverted pendulum \cite{LL}:  if the point of support of a plane 
pendulum in a uniform gravitational field is subject to rapid oscillations, a 
new minimum in the effective potential energy can be induced.  The quantum
fluctuations of the dot population resemble oscillations of the support, and 
the mechanical analog of the SQUID (and of the Coulomb blockade problem) is 
a slight modification of the simple pendulum \cite{SZ}.  The contribution
to the pendulum's potential energy from gravity 
is given by the second term of (\ref{sgclenergy}) with $2\pi\phi$ analogous 
to the pendulum's angular position.  The first term of (\ref{sgclenergy}) 
describes a harmonic torsion bar one end of which is attached directly to the 
pendulum axle while the other is held at some fixed angle, this angle
being analogous to the gate voltage, $n_{x}$, of (\ref{sgclenergy}). The
inverted pendulum analogy is at best suggestive, and our effect is weaker -
we only find a flattening of the maxima of the periodic part of the dot energy.

Direct experimental observation of the intermediate phase with
integer-half-integer sequences of plateaus might encounter several
difficulties.  The 
fluctuations around the half-integer dot population are large and likely
a model-independent effect.  Moreover in the regime of parameters, 
$b \ll 1$ and $g \le 1$, the range of existence of the intermediate phase is 
extremely narrow.  

If the plateaus were to be observed, the corresponding experimental
signal of entering this phase from the continuum phase would be a divergence 
of the gate-reservoir capacitance (proportional to $dn/dn_{x}$) at the lower 
critical value of the amplitude of the periodic potential, 
$v_{c1}$, at two values of the gate voltage placed 
symmetrically about $n_{x} = 1/2$ (for $g = 1$ these are 
$n_{x} \approx 0.3771$ and $n_{x} \approx 0.6229$).  Inside the intermediate 
phase these two peaks would move towards each other (and $n_{x} = 1/2$) as the 
dot is made more isolated, eventually coalescing as $v$ approaches $v_{c2}$.  
An alternative way of observing the existence of the half-integer 
(and also integer) plateaus is through detecting the associated hysteresis 
phenomena which must accompany the first-order population jumps.  We believe 
that the observation of hysteresis phenomena is a good gauge to distinguish 
between the predicted continuous and discontinuous $n(n_{x})$ dependencies.

We believe, additionally, that these plateaus should be accessible numerically.
Monte Carlo simulations of actions related to that found in (\ref{Action}) 
have already been successively accomplished.  Specifically setting $b = 0$ in 
(\ref{sgclenergy}) gives us an action appropriate to describing tunneling 
between quantum Hall edges.  Monte Carlo simulations of 
this action were done in \cite{MC} which later were shown to match exact 
scaling curves \cite{FLS}.  Moreover very accurate Monte Carlo simulation of 
the classical one-dimensional Ising model with interactions decaying as 
inverse-square of distance was recently completed \cite{LM}, and the results 
were found to be in very good agreement with theoretical predictions 
\cite{Anderson&Yuval,Bray&Moore}.  We are thus currently examining numerically 
the possibility of observing the half-integer plateaus \cite{RE}.

Having outlined the prospects of observing the intermediate phase, 
we may note that in the context of adsorption phenomena the existence of a 
regime with stable fractional film thickness centered about half-layer 
position is experimentally well-established \cite{Hess}.  In fact, the 
effects observed are even more puzzling - in a range of parameters variations 
in the external pressure lead to film thickness jumps {\it only} between 
half-integer layer coverages.  The regime analogous to our 
finding - integer-half-integer adsorption curve was discussed theoretically 
\cite{Weichman&Prasad}, but not yet seen experimentally.

We would also like to mention a recent Little-Parks type experiment 
on small disordered $AuZn$ cylindrical films where a resistance oscillation 
with a period of $\Phi_{0}/2$, half of the flux quantum, was observed 
\cite{Zadorozhny}.  The new resistance minima develop below certain 
temperatures and coexist with the standard Little-Parks minima corresponding 
to the integer number of flux quanta.  This closely resembles our prediction 
for the intermediate phase translated into the superconductivity language.    

The persistence of the half-integer minimum in the population dependence of 
the ground-state energy at the charge degeneracy points is the ultimate reason
why the physics of the Kondo/Ising systems is different from that of the
tricritical systems.  The presence of an extra minimum in the latter case 
makes it more difficult for the electrons to enter the dot, thus enhancing 
the effects of Coulomb blockade.

Putting aside the existence of the intermediate phase, in the $g < 1$ regime 
there is nonetheless a qualitative correspondence between the results of 
Sections III and IV - for fixed $g$ and sufficiently small $v$ 
(sufficiently large bare tunneling amplitude $\Delta_{0}$) the $n(n_{x})$ 
dependence is continuous while above a critical $v$ (below critical 
$\Delta_{0}$) the $n(n_{x})$ dependence is discontinuous - it is a modified
staircase with plateaus centered around integer dot population.  The 
correspondence can be made more quantitative by comparing the dot population 
jump between two neighboring integer plateaus numerically evaluated at the 
upper critical curve $v_{c2}$ from (\ref{varenergy}) - (\ref{neq}) with the 
universal prediction $\Delta n = g^{1/2}$ for the two-state system (i.e. a
ferromagnetic Ising model with interactions decaying as inverse square of 
distance).  We found that whenever the width of the intermediate phase is
small, i.e. for (i) $0 < g < 1$ and $b \rightarrow 0$ or (ii) 
$g \rightarrow 0$ and arbitrary $b$, the population jump computed from 
(\ref{varenergy}) - (\ref{neq}) is well approximated by $g^{1/2}$.  For 
$g \rightarrow 0$, a little algebra shows that $\Delta n \sim g^{1/2}$ is
a consequence of Eqs. (\ref{meq}) and (\ref{neq}), and the
prefactor can be established numerically.  We find that  
$\Delta n = 1.01991g^{1/2}$ 
as $g \rightarrow 0$ which is very close to the theoretical expectation 
$\Delta n = g^{1/2}$.  

For the experimentally most significant case when the electrons in the 
reservoirs are noninteracting, $g = 1$, the discrepancy between the 
systems of the Kondo/Ising and the tricritical subclasses is largest.  In the 
former case a continuous $n(n_{x})$ dependence is predicted for arbitrarily 
small bare tunneling amplitude, while in the latter case, a continuous, 
intermediate staircase, and modified integer staircase are predicted, 
as in Fig.2, as the tunneling amplitude decreases ($v$ increases).  
This difference between the two types of behavior is unsurprising as the 
treatment of Section III for the case $g = 1$ was already extremely 
delicate - the slightest change in the physics such as a stronger suppression 
of the number fluctuations on the dot might change the outcome qualitatively.  
Experimental or numerical observation for $g = 1$ of the modified Coulomb 
staircase with first-order jumps between integer population plateaus will 
provide indirect evidence for the intermediate phase -- they either exist 
together (tricritical type) or do not exist at all (Kondo/Ising type).  It 
would be also interesting to verify the universality of the $n(n_{x})$ 
dependence (\ref{nsol}) by varying the size and shape of large dots.

Beyond the existence of the half-integer plateaus, an important issue
arises in relating the smearing of the Coulomb staircase found
in \cite{Matveev91} with our results.  One possible answer is that the
effective action implicit in \cite{Matveev91} belongs to
what we call the Kondo/Ising subclass (specifically
a Kondo Hamiltonian with the coupling $J_z$ set to 0)
while our variational finding of a staircase
occurs solely for actions falling in the tricritical subclass.  In 
Section III we demonstrated that after the dot problem is truncated to 
a two-state system, {\it any} model with an effective
free-fermion reservoir will predict 
a smeared Coulomb staircase.  Because in the classical limit of a
closed dot 
the free-fermion case is marginal, it is also important to critically 
reexamine 
the validity of the two-state approximation.  Let us assume that the
gate voltage is tuned at a charge degeneracy point, and that the full bare 
potential (\ref{Clenergy}) has metastable minima ignored in the 
two-state approximation.  Because these extra minima can now be populated with 
small finite probability, the full potential (\ref{Clenergy}) 
suppresses
electron number fluctuations on the dot somewhat {\it stronger} than its 
two-state truncation 
discussed in Section III \cite{note}.  Whether this suppression is 
strong enough to restore a true staircase, remains an open problem.

We have also conducted a study of the spinful problem akin to the variational 
analysis of Section IV.  The results will be presented elsewhere.  Similar to 
the spinless case, the spin-1/2 Coulomb blockade problem is dual to that of a 
two-junction SQUID \cite{Likharev}.  Although this problem is more involved, 
for the case when the electrons are noninteracting, the model is a natural
generalization of (\ref{sgclenergy}).  We find that for sufficiently 
small transparency of the constriction there are two staircase phases in close
correspondence with the spinless case.  Experimentally \cite{Berman} for 
spinful electrons, the dot population was found to change continuously with 
the gate voltage for a nearly open contact while developing smeared step-like 
structures with the contact nearly closed.  The shape of the smeared step for 
the almost closed dot was found to be consistent with a finite temperature 
generalization of the perturbation theory done in \cite{Matveev91}.  However 
the temperature of the dot was far above the Kondo temperature found
in \cite{Matveev91}, the scale governing the putative smearing of the
steps at zero temperature.  Whether the steps remain smeared at $T=0$ is thus 
an unanswered experimental question.  

We hope that experimental attempts will be made in the near future to test 
our predictions.
         
\section{ACKNOWLEDGMENTS}  

We thank B. S. Deaver, P. Fendley, M. Fowler, M. A. Moore, and 
T. J. Newman for valuable discussions.  R.K. acknowledges helpful
conversations with Shivaji Sondhi, Matthew Fisher, Vadim Oganesyan,
and Karyn Le Hur.

E.B.K. and X.Q. would like to thank both the Thomas F. Jeffress and Kate 
Miller Jeffress Memorial Trust, and the Chemical Sciences, Geosciences and 
Biosciences Division, Office of Basic Energy Sciences, Office of Science, U. S.
Department of Energy.  R. K. thanks the support of the NSF through grant 
DMR-9802813.


\begin{references}
\bibitem{next}  The parameter $n_{x}$ is an ``external'' number of 
electrons imposed by the dot environment; it is generally different from the 
actual population of the dot $n$.
\end{references}
\end{document}